\documentclass[11pt,a4paper]{article}
\usepackage{amsmath}
\usepackage{amsfonts, graphics}
\usepackage[T1]{fontenc}
\usepackage[utf8]{inputenc}
\usepackage{authblk}

\title{A Model of Emergent Universe in Inhomogeneous Space-Time}
\author[1]{Subhra Bhattacharya\thanks{subhra.maths@presiuniv.ac.in}}
\author[2]{S. Chakraborty\thanks{schakraborty.math@gmail.com}}
\affil[1]{Department of Mathematics, Presidency University, Kolkata 700 073, INDIA}
\affil[2]{Department of Mathematics, Jadavpur University, Kolkata 700 032, INDIA}

\begin{document}
  \maketitle

\newcommand{\IC}{\mathbb C}
\newcommand{\IR}{\mathbb R}
\newcommand{\IQ}{\mathbb Q}
\newcommand{\IZ}{\mathbb Z}
\newcommand{\IN}{\mathbb N}
\newcommand{\mk}{\marginpar}
\newcommand{\Rn}{\IR^n}
\newcommand{\be}{\begin{enumerate}}
\newcommand{\ee}{\end{enumerate}}
\newcommand{\dl}{\displaystyle}

\hrulefill
\vspace{-.5em}

\vspace{1em}

{\large\bf Abstract:}
\vspace{1em}

A scenario of an emergent universe is constructed in the background of an inhomogeneous space-time model which is asymptotically (at spatial infinity) FRW space-time. The cosmic substratum consists of non-interacting two components, namely {\bf a)} homogeneous and isotropic fluid but dissipative in nature and {\bf  b)} an inhomogeneous and anisotropic barotropic fluid. In non-equilibrium thermodynamic prescription (second order deviations), particle creation mechanism is considered the cause for the dissipative phenomena. It is found that for constant value of the particle creation rate parameter there exists a scenario of emergent universe.

\vspace{1em}

{\bf Keyword:}

\vspace{1em}

\texttt{Emergent universe, inhomogeneous space-time, particle creation}

\hrulefill
\vspace{-.5em}

\vspace{1em}

\vspace{4em}

There are lot of works to remove the bad feature of standard cosmology i.e. the initial big bang singularity. Usually, the different cosmological scenarios proposed to overcome this non-physical aspect are classified into two groups: {\bf i)} model for bouncing universes or {\bf ii)} model of emergent universes. The second alternative is related to singularity free inflationary model in context of classical general relativity and its formulation is the aim of the present work. The scenario of emergent universe has the following features: {\bf a)} ever existing, {\bf b)} absence of any time-like singularity and {\bf c)} almost static behaviour in the infinite past ($t\rightarrow-\infty$) ( and then evolves into an inflationary stage). Also in the literature, the model of emergent universe is considered as a modern and extended version of the Lemaitre-Eddington universe.

The proposal for emergent scenario was started long back in 1967 by {\it Harrison} [1]. He obtained a closed model of the universe with radiation which in the asymptotic past is nothing but the state of an Einstein static model. Then after a gap of nearly forty years, {\it Ellis and Maartens} [2], {\it Ellis et.al.} [3] formulated a model of closed universe with non-interacting two fluid components: a minimally coupled scalar field with a typical self interacting potential and a barotropic fluid having equation of state: $p=\omega \rho~(-\frac{1}{3}\leq\omega\leq 1).$ Although they were not able to give analytic solution, only the asymptotic behaviour had the characteristic of emergent universe. Then, using Starobinsky model {\it Mukherjee et. al.} [4] formulated solutions having emergent universe scenario. Subsequently Mukherjee and collaborators [5] presented a general framework for an emergent universe model within adhoc equation of state, having exotic behaviour in some phases. Since then there are series of works [6-18] either in different gravity theories or with different types of matter or both with a view to propose a model for emergent universe scenario. Recently, an emergent universe model has been proposed [19] where the initial static state is characterized by a scalar field in false vacuum which decays to a state of true vacuum through quantum tunnelling. Very recently, the idea of emergent universe has been constructed using the idea of particle creation in the perspective of non-equilibrium thermodynamics both in Einstein gravity [20] as well as in brane model [21] (RS II and DGP). Lastly, about few months earlier, {\it Paul et. al.} [22] investigated the emergent universe scenario in the presence of interacting (two or three) fluids having non-linear equation of state.

The present work is an extension (in some sense) of the studies in references [20] and [21]. The universe is chosen as inhomogeneous spherically symmetric space-time which is asymptotically (at spatial infinity) a FRW model. This typical inhomogeneous space-time was studied by {\it Cataldo et. al.} [23] in studying evolving Lorentzian wormholes supported by phantom matter with constant state parameters. Non-interacting two-fluid system is the matter content of the system. One of the fluids is homogeneous and isotropic but dissipative in nature, while the other matter component is an inhomogeneous and anisotropic barotropic fluid. The particle creation mechanism in the non-equilibrium thermodynamic prescription is assumed to be the cause of the dissipative phenomena. The following two things motivate us in the present work. Firstly it is interesting to have an initial singularity (big-bang) free solution for inhomogeneous model. Secondly, at very early phase of its evolution, the universe was very likely to be in a state of much disorder, that is inhomogeneity was important at that era, so it is interesting to have emergent scenario with inhomogeneous space time.
\vspace{4em}

The line element for inhomogeneous spherically symmetric space-time is given by [23]
 
\begin{equation}
ds^{2}=-dt^{2}+a^{2}(t)\left[\frac{dr^{2}}{1-b(r)}+r^{2}d\Omega^{2}\right].
\end{equation}

 The energy momentum tensor for fluid I, which is homogeneous and isotropic but dissipative in character is given by
 \begin{equation}
  T_{\mu\nu}^{(I)}=(\rho_{1}+p_{1}+\Pi_{1})u_{\mu}u_{\nu}+(p_{1}+\Pi_{1})g_{\mu\nu}
\end{equation} 
 where $\rho_{1}=\rho_{1}(t),~p_{1}=p_{1}(t)$ and $\Pi_{1}=\Pi_{1}(t)$ are respectively the energy density, isotropic pressure and the pressure due to dissipation.
 
 For the fluid II which is both inhomogeneous and anisotropic in nature, the energy-momentum tensor has the expression
 \begin{equation}
 T_{\mu\nu}^{(II)}=(\rho_{2}+p_{2t})v_{\mu}v_{\nu}+p_{2t}g_{\mu\nu}+(p_{2r}-p_{2t})x_{\mu}x_{\nu}
 \end{equation}
where $\rho_{2}=\rho_{2}(t,r),~p_{2r}=p_{2r}(t,r)$ and $p_{2t}=p_{2t}(t,r)$ are respectively the energy density, radial and transverse pressures of fluid II. Here $v_{\mu}$ and $x_{\mu}$ are unit time-like and space-like vectors respectively, satisfying
\begin{equation}
v_{\mu}v^{\mu}=-x_{\mu}x^{\mu}=-1,~x^{\mu}v_{\mu}=0.
\end{equation}

As the fluid components are non-interacting so the separate conservation equations for both the fluids are
 \begin{eqnarray}
 \frac{\partial\rho_{1}}{\partial t}+3H(\rho_{1}+p_{1}+\Pi_{1})=0 \label{econsv1}\\
 \frac{\partial\rho_{2}}{\partial t}+H(3\rho_{2}+p_{2r}+2p_{2t})=0\label{econsv2}\\
 \frac{\partial p_{2r}}{\partial r}=\frac{2}{r}(p_{2t}-p_{2r})\label{econsv3}
 \end{eqnarray}
 Note that the equations (\ref{econsv1}) and (\ref{econsv2}) are the usual conservation equations for the two fluids while equation (\ref{econsv3}) can be termed as relativistic Euler equation.
 
 For the present model, the explicit form of the Einstein field equations are [24-26]
 \begin{eqnarray}
 3H^{2}+\frac{b+rb^{'}}{a^{2}r^{2}}=\kappa\rho_{1}+\kappa\rho_{2}\label{field1}\\
 -(2\dot{H}+3H^{2})-\frac{b}{a^{2}r^{2}}=\kappa(p_{1}+\Pi_{1})+\kappa p_{2r}\label{field2}\\
 -(2\dot{H}+3H^{2})-\frac{b^{'}}{2a^{2}r}=\kappa(p_{1}+\Pi_{1})+\kappa p_{2t}\label{field3}
 \end{eqnarray}
where dot indicates differentiation with respect to $t$ and prime indicates differentiation with respect to $r.$ To have an explicit solution of the above Einstein field equations we assume that for fluid II, the anisotropic pressure components satisfy barotropic equations of state, namely,
\begin{equation}
p_{2r}=\omega_{r}\rho_{2} ~\text{and}~ p_{2t}=\omega_{t}\rho_{2}\label{eqs}
\end{equation}
 where the equation of state parameters $\omega_{r}$ and $\omega_{t}$ are assumed to be constant. Now solving the conservation equations (\ref{econsv2}) and (\ref{econsv3}) the energy density of fluid II is obtained as
 \begin{equation}
 \rho_{2}(t,r)=\rho_{0}\frac{r^\frac{2(\omega_{t}-\omega_{r})}{\omega_{r}}}{a^{3+\omega_{r}+2\omega_{t}}}\label{ed1}
 \end{equation}
 where $\rho_{0}$ is the constant of integration.
 
 The function $b(r)$ can be solved by comparing the field equations (\ref{field2}) and (\ref{field3}) and using equations (\ref{eqs}) and (\ref{ed1}), [23] as
 \begin{equation}
 b(r)=b_{0}r^{2}-\kappa\rho_{0}\omega_{r}r^{-(\frac{1}{\omega_{r}}+1)},\label{b}
 \end{equation}
 ($b_{0}$ being the constant of integration. It should be noted here that $b_{0}$ actually corresponds to the curvature constant $k$ ) provided $\omega_{r}$ and $\omega_{t}$ are restricted by the relation [23]
 \begin{equation}
 \omega_{r}+2\omega_{t}+1=0\label{const1}.
 \end{equation}
 So eliminating $\omega_{t}$ between (\ref{ed1}) and (\ref{const1}) we have
 \begin{equation}
 \rho_{2}(t,r)=\frac{\rho_{0}}{a^{2}}r^{-(3+\frac{1}{\omega_{r}})}\label{rho}
 \end{equation}
 with \begin{equation}
 p_{2r}=\omega_{r}\rho_{2}~\text{and} ~p_{2t}=-\frac{1+\omega_{r}}{2}\rho_{2}.\label{p}
 \end{equation}
Using the fluid II components from equations (\ref{rho}) and (\ref{p}) and also using equation (\ref{b}) for $b(r),$ the field equations (\ref{field1})-(\ref{field3}) simplifies to  
\begin{eqnarray}
3\left(H^{2}+\frac{b_{0}}{a^{2}}\right)=\kappa\rho_{1}\label{field4}\\
-\left(2\dot{H}+3H^{2}+\frac{b_{0}}{a^{2}}\right)=\kappa(p_{1}+\Pi_{1})\label{field5}
\end{eqnarray}
which are nothing but Einstein field equations (i.e. Friedman equations) for homogeneous and isotropic space-time model, with curvature constant equivalent to $b_{0}$. Note that the energy density $\rho_{1}$ and fluid pressure $p_{1}$ are related by the conservation equation (\ref{econsv1}).

As we have mentioned earlier that the present work is related to particle creation mechanism (for an open thermodynamical system) in non-equilibrium thermodynamical aspect, so the non-conservation of particle number is characterised by [27-30]
\begin{equation}
\dot{n}+\theta n=n\Gamma\label{nconsv}
\end{equation}
where $n=\frac{N}{V}$ is the particle number density and $N$ is the number of particles in a co-moving volume $V.$ The scalar $\theta=u^{\mu}_{;\mu}$ represents the fluid expansion, particle flow vector is defined by $N^{\mu}=nu^{\mu},$ $\Gamma$ represents the particle creation rate and $\dot{n}$ by notation represents $n_{,\mu}u^{\mu}.$ It should be noted that there is some dissipation effect due to particle creation and effectively there is non-equillibrium thermodynamics [30]. Also, particle creation $(\Gamma>0)$ or annihilation $\Gamma<0)$ is characterised by the sign of $\Gamma.$

 We shall now relate the particle creation parameter $\Gamma$ to the dissipative pressure $\Pi_{1}$ of the present work. The Gibb's equation can be written as [27-30]
\begin{equation}
Tds=d\left(\frac{\rho_{1}}{n}\right)+p_{1}d\left(\frac{1}{n}\right)\label{gibb}
\end{equation} 
 where $s$ is the entropy per particle, $T$ is the fluid temperature and Clausius relation is taken care of. Using conservation equations (\ref{econsv1}) and (\ref{nconsv}), a simple algebra gives the entropy variation from the above Gibb's equation, as [29-31]
 \begin{equation}
 nT\dot{s}=-\Pi_{1}\theta-\Gamma(p_{1}+\rho_{1}).\label{gibbs1}
 \end{equation}
In particular if we are restricted to adiabatic (i.e. isentropic) thermal process (for which entropy per particle is constant, i.e. $\dot{s}=0$) then from the above relation (\ref{gibbs1}) we have
\begin{equation}
\Pi_{1}=-\frac{\Gamma(p_{1}+\rho_{1})}{\theta}.\label{pi}
\end{equation} 
It shows that for isentropic thermal process dissipative pressure is completely determined by particle creation rate. Equivalently, the physical process may be considered as a perfect fluid with barotropic equation of state $p_{1}=\omega_{1}\rho_{1}$ and particle creation mechanism causes the dissipative phenomena. In this context one should keep in mind, that in adiabatic system entropy production is caused by particle creation and by enlargement of the phase space due to expansion of the universe. Thus the present non-equilibrium configuration is not the usual one, rather a state with equilibrium properties as well. Now eliminating $\Pi_{1},$ from equation (\ref{pi}) and the field equation (\ref{field5}) (for simplicity we assumed a flat model for space-time and hence $b_{0}$ is assumed to be zero) and using the equation of state parameter $\omega_{1}=\frac{p_{1}}{\rho_{1}},$ we find that the particle creation parameter is related to the evolution of the universe as 
\begin{equation}
\frac{\Gamma}{3H}=1+\frac{2}{3(1+\omega_{1})}\frac{\dot{H}}{H^{2}.}\label{de1}
\end{equation}

Now considering the above equation (\ref{de1}) as the evolution equation for Hubble parameter we have on integration 
\begin{equation}
H=\frac{\exp[\frac{1+\omega_{1}}{2}\int\Gamma dt]}{H_{0}+\frac{3(1+\omega_{1})}{2}\int \exp\{\frac{1+\omega_{1}}{2}\int\Gamma dt\}dt}\label{H}
\end{equation}
and consequently
\begin{equation}
a=a_{0}\left[H_{0}+\frac{3(1+\omega_{1})}{2}\int\exp\left\{\left(\frac{1+\omega_{1}}{2}\right)\int\Gamma dt\right\}dt\right]^{2/3(1+\omega_{1})}\label{a}
\end{equation}
where $H_{0}$ and $a_{0}$ are constants of integration.

In particular for constant particle creation rate, i.e. $\Gamma=\Gamma_{0}$ the above solution simplifies to 
\begin{eqnarray}
H=\exp\left\{\frac{1+\omega_{1}}{2}\Gamma_{0}t\right\}\left[H_{0}+\frac{3}{\Gamma_{0}} \exp\left\{\frac{1+\omega_{1}}{2}\Gamma_{0}t\right\}\right]^{-1}\label{FH}\\
a=a_{0}\left[H_{0}+\frac{3}{\Gamma_{0}}\exp\left\{\left(\frac{1+\omega_{1}}{2}\right)\Gamma_{0}t\right\}\right]^{2/3(1+\omega_{1})}\label{Fa}\\
\rho_{1}(t)=3\exp\{(1+\omega_{1})\Gamma_{0}t\}\left[H_{0}+\frac{3}{\Gamma_{0}}\exp\left\{\left(\frac{1+\omega_{1}}{2}\right)\Gamma_{0}t\right\}\right]^{-2}\label{rho1}\\
\rho_{2}(t,r)=\frac{\rho_{0}}{a_{0}^{2}}r^{-(3+\frac{1}{\omega_{r}})}\left[H_{0}+\frac{3}{\Gamma_{0}}\exp\left\{\left(\frac{1+\omega_{1}}{2}\right)\Gamma_{0}t\right\}\right]^{-4/3(1+\omega_{1})}.\label{rho2}
\end{eqnarray}
 
The solution has the following features in the asymptotic past: (assuming $1+\omega_{1}>0$)
\begin{align}
\text{i)}~ &H\rightarrow 0,~a\rightarrow (H_{0})^{\frac{2}{3(1+\omega_{1})}}a_{0}=a_{i}~ \text{as} ~t\rightarrow -\infty \nonumber \\
\text{ii)}~ &H\simeq 0,~ a\simeq a_{i}~\text{for}~ t\ll 0\\
\text{iii)}~&H\simeq\frac{\Gamma_{0}}{3},~a\simeq a_{0}\left(\frac{3}{\Gamma_{0}}\right)^{2/3(1+\omega_{1})}\exp\left(\frac{1}{3}\Gamma_{0}t\right),~ \text{for}~ t\gg 0.\nonumber
\end{align}

Hence the above cosmological solution describes a scenario of emergent universe. Also in the above asymptotic limit the energy densities of the two fluid components take the form:
\begin{align}
\text{i)}~&\rho_{1}\rightarrow 0, ~\rho_{2}\rightarrow \frac{\rho_{0}r^{-(3+1/\omega_{r})}}{a_{0}^{2}(H_{0})^{4/3(1+\omega_{1})}}=\rho_{i}r^{-(3+1/\omega_{r})}~\text{as}~t\rightarrow -\infty\nonumber
\\
\text{ii)}&~\rho_{1}\simeq 0,~\rho_{2}\simeq\rho_{i}r^{-(3+1/\omega_{r})}~\text{for}~t\ll 0\\
\text{iii)}&~\rho_{1}\simeq \frac{\Gamma_{0}^{2}}{3},~\rho_{2}\simeq \rho_{0}\left(\frac{\Gamma_{0}}{3}\right)^{4/3(1+\omega_{r})}r^{-(3+1/\omega_{r})}\exp\left(-\frac{2}{3}\Gamma_{0}t\right) ~\text{for}~ t\gg 0 \nonumber.
\end{align}

Also for equation of state parameter $\omega_{r}>0$ at spatial infinity (i.e. $r\rightarrow \infty$) the present inhomogeneous model describing a flat FRW universe, as expected gives $\rho_{2}\rightarrow 0.$ Thus in the spatial asymptotic limit the present inhomogeneous model becomes homogeneous and isotropic FRW universe supported by a single perfect fluid with dissipative effect. It is worthy to mention that for such model, emergent scenario has already been discussed in reference [20]. Normally, in emergent models of universe, there is a phase of Einstein static model in infinite past followed by a period of inflationary expansion. Usually inflationary phase is described by a scalar field (inflaton) having self interacting potential. However, in the present problem, if the homogeneous fluid is not phantom in nature (that is $1+\omega_{1}>0$) then the universe is always in a state of super-inflation (i.e. $\dot{H}>0$). Further, for $t=-\epsilon$ with $|\epsilon|<<1$ we have $$H\simeq\frac{\Gamma_{0}}{3},~a\simeq a_{0}\left(\frac{3}{\Gamma_{0}}\right)^{\frac{2}{3(1+\omega_{1})}}\exp\left(\frac{\Gamma_{0}}{3}t\right),$$ a phase of exponential expansion (inflationary era) without any inflaton. Particle creation mechanism drives the inflationary epoch. Therefore, a scenario of emergent universe is possible for inhomogeneous  space-time model. Finally as a remark we mention that the present model has the basic features of steady state theory of {\it Fred Hoyle et.al} [32, 33] and closely resembles with it.

\vspace{1em}
{\large\bf Acknowledgements:}
\vspace{1em}

SB is thankful to UGC for financial support under their Faculty Recharge Program.
SC thanks IUCAA for their research facility.

\vspace{1em}
{\large\bf References:}
\vspace{1em}

\begin{description}
\item[[1]] E. R. Harrison, Mon. Not. R. Astron. Soc. 137 (1967) 69.
\item[[2]] G. F. R. Ellis, R. Maartens, Class. Quant. Grav.  21 (2004) 223, arXiv:gr-qc/0211082.
\item[[3]] G. F. R. Ellis, J. Murugan, C. G. Tsagas, Class. Quant. Grav. 21 (2004) 233, arXiv: gr-qc/0307112.
\item[[4]] S. Mukherjee, B. C. Paul, S. D. Maharaj, A. Beesham, (2005), arXiv: 0505103 [gr-qc].
\item[[5]] S. Mukherjee, B. C. Paul, N. K. Dadhich, S. D. Maharaj, A. Beesham, Class. Quant. Grav. 23 (2006) 6927, arXiv: gr-qc/0605134.
\item[[6]] D. J. Mulryne, R. Tavakol, J. E. Lidsey, G. F. R. Ellis, Phys. Rev. D 71 (2005) 123512, arXiv: astro-ph/0502589.
\item[[7]] A. Banerjee, T. Bandyopadhyay, S. Chakraborty, Gravitation and Cosmology 13 (2007) 290, arXiv: 0705.3933 [gr-qc];

A. Banerjee, T. Bandyopadhyay, S. Chakraborty, Gen. Relt. Grav. 40 (2008) 1603, arXiv: 0711.4188 [gr-qc].
\item[[8]] N. J. Nunes, Phys. Rev. D 72 (2005) 103510, arXiv: astro-ph/0507683.
\item[[9]] J. E. Lidsey, D. J. Mulryne, Phys. Rev. D 73 (2006) 083508, arXiv: hep-th/0601203.
\item[[10]] U. Debnath, Class. Quant. Grav. 25 (2008) 205019, arXiv: 0808:2379 [gr-qc].
\item[[11]] B. C. Paul, S. Ghose, Gen. Relt. Grav.  42 (2010) 795, arXiv: 0809.4131 [hep-th].
\item[[12]] U. Debnath, S. Chakraborty, Int. J. Theo. Phys. 50 (2011) 2892, arXiv: 1104.1673 [physics.gen-ph].
\item[[13]] S. Mukerji, N. Mazumder, R. Biswas, S. Chakraborty, Int. J. Theo. Phys. 50 (2011) 2708, arXiv: 1106.1743 [gr-qc].
\item[[14]]K. Zhang, P. Wu, H. Wu, JCAP. 01 (2014) 048. 
\item[[15]] A. Beesham, S. V. Chervon, S. D. Maharaj, A. S. Kubasov Quantum Matter 388 (2013) 2. 
\item[[16]] R. Ghosh, S. Chattopadhyay, Eur. Phys. J. Plus 128 (2013) 12. 
\item[[17]] S. Ghose, P. Thakur, B. C. Paul, Mon. Noc. Roy. Astron. Soc. 421 (2012) 20. 
\item[[18]] S. Ghose, B. C. Paul, arXiv: 1102.3539.
\item[[19]] P. Labrana, Phys. Rev. D 86 (2012) 083524, arXiv:1111.5360 [gr-qc].
\item[[20]] S. Chakraborty, Phys. Lett. B 732 (2014) 81, arXiv: 1403.5980 [gr-qc].
\item[[21]] J. Dutta, S. Haldar, S. Chakraborty, arXiv: 1505.01263 [gr-qc].
\item[[22]] B. C. Paul, A. Majumdar, Class. Quant. Grav. 32 (2015) 115001, arXiv: 1503.08248 [gr-qc]
\item[[23]] M. Cataldo, P. Labrana, S. del Campo, J. Crisostomo, P. Salgado, Phys. Rev. D 78 (2008) 104006, arXiv: 0810.2715 [gr-qc].
\item[[24]] M. Cataldo, S. del Campo, Phys. Rev. D 85 (2012) 104010, arXiv: 1204.0753 [gr-qc].
\item[[25]] M. Cataldo, P. Mella, P. Minning and J. Saavedra, Phys. Lett. B 662 (2008) 314, arXiv: 0803.1086 [hep-th].
\item[[26]] S. Pan, S. Chakraborty, Eur. Phys. J. C, 75, (2015), 21, arXiv: 1412.6094 [gr-qc].
\item[[27]] T. Harko, F. S. N. Lobo, Phys. Rev. D 87 (2013) 044018, arXiv: 1210.3617 [gr-qc].
\item[[28]] S. Pan, S. Chakraborty, Adv. High Energy Phys. 2015 (2015) 654025, arXiv: 1404.3273 [gr-qc].
\item[[29]] S. Chakraborty, S. Saha, arXiv: 1402.6944 [gr-qc].
\item[[30]] S. Chakraborty, S. Saha, Phys. Rev. D 90 (2014) 123505.
\item[[31]] W. Zimdahl, Phys. Rev. D 61 (2000) 083511, arXiv: astro-ph/9910483.
\item[[32]] F. Hoyle, G. Burbidge, J. V. Narlikar, Astrophysical Journal 410 (1993) 437;

F. Hoyle, G. Burbidge, J. V. Narlikar, Mon. Noc. Roy. Astron. Soc. 267 (1994) 1007;

F. Hoyle, G. Burbidge, J. V. Narlikar, Mon. Noc. Roy. Astron. Soc. 269 (1994) 1152;

F. Hoyle, G. Burbidge, J. V. Narlikar, Proc. Roy. Soc. A 448 (1995) 191.
\item[[33]] F. Hoyle, G. Burbidge, J. V. Narlikar, A Different Approach to Cosmology, Cambridge University Press, 2000.

\end{description}
\end{document}